# Robustness of Greenberger-Horne-Zeilinger and *W* states for teleportation in external environments

Ming-Liang Hu[*]

*School of Science, Xi'an University of Posts and Telecommunications, Xi'an 710061, China*

**Abstract:** By solving analytically a master equation in the Lindblad form, we study quantum teleportation of the one-qubit state under the influence of different surrounding environments, and compared the robustness between Greenberger-Horne-Zeilinger (GHZ) and *W* states in terms of their teleportation capacity. The results revealed that when subject to zero temperature environment, the GHZ state is always more robust than the *W* state, while the reverse situation occurs when the channel is subject to infinite temperature or dephasing environment.

**Keywords:** Teleportation, Decoherence

**PACS number(s):** 03.67.Hk, 03.65.Yz, 75.10.Pq

Starting from the seminal protocol formulated in Ref. [1] an increasing attention has been devoted to the investigation of quantum teleportation both at theoretical [2–7] and experimental level [8–11]. Differently from its classical counterpart [12], quantum teleportation enables the reconstruction of an arbitrary unknown quantum state at a spatially distant location with unity fidelity under the help of classical communication and without the need of transferring any particles physically. The key requirements for this quantum protocol are the performance of clean projective measurements and the prior shared maximally entangled channel state between the sender Alice and the recipient Bob [1–5]. However, in real circumstances it is usually very difficult to prepare maximally entangled resource because during the preparation and distribution process the decoherence may take effect due to the unavoidable interaction of the system with its surrounding world [2,13]. Moreover, while Alice and Bob perform the Bell basis measurements and the conditional unitary operations the decoherence may also be set in [14,15]. All these can play a significant role in reducing the fidelity of the expected outcomes, and we shall discuss one such situation later in this Letter.

The original protocol of teleportation proposed by Bennett et al. [1] is implemented through a channel involving an Einstein-Podolsky-Rosen (EPR) pair previously shared by Alice and Bob, and this is the most economical resource for teleporting the one-qubit state since for teleporting an arbitrary *N*-qubit state perfectly the channel should possess at least 2*N* qubits [1,4,5]. After the milestone work of Bennett et al. [1], teleportation through quantum channels contain more than two qubits has also been discussed extensively. In particular, in Refs. [16,17] the authors found that the three-partite Greenberger-Horne-Zeilinger (GHZ) state and the *W* state can also be used as quantum channels for perfect one-qubit teleportation. As we know, the GHZ and the *W* states are two inequivalent classes of multipartite entangled states under stochastic local operations and classical communication (SLOCC) [18]. While they both allow perfect one-qubit teleportation in the idealistic situation, a question naturally arises at this stage is which of them is more robust when being used as quantum channels for teleportation in real world? In a recent work [19] Jung et al. studied such a problem by introducing noises into the teleportation process, and they found that the answer to this question is dependent on the type of noisy channel.

---
[*] Corresponding author.
Tel.: +86 029 88166094
*E-mail address*: mingliang0301@163.com, mingliang0301@xupt.edu.cn (M.-L. Hu)

In this Letter we investigate possibility of quantum teleportation of the one-qubit state for the situation in which the system is subject to different sources of decoherence. We will concentrate our attention to the comparison of the GHZ and *W* states in terms of their robustness as quantum channels for teleportation in zero temperature environment, infinite temperature environment and dephasing environment [13], for which the decoherence dynamics of the system can be described by a general master equation in the Lindblad form [20,21]

$$\frac{d\rho}{dt} = \frac{\gamma}{2} \sum_{k,i} \left( 2 L_{k,i} \rho L_{k,i}^\dagger - L_{k,i}^\dagger L_{k,i} \rho - \rho L_{k,i}^\dagger L_{k,i} \right), \tag{1}$$

where $\rho$ and $\gamma$ denote, respectively, the reduced density operator of the system and the coupling strengths of the qubits with their respective environment. In contrast to Refs. [14,19], the Lindblad operators here are defined in terms of the raising and lowering operators $\sigma^\pm = (\sigma^x \pm i\sigma^y)/2$ as $L_k = \sigma_k^-$ for zero temperature environment, $L_{k,1} = \sigma_k^-$ and $L_{k,2} = \sigma_k^+$ for infinite temperature environment, and $L_k = \sigma_k^+ \sigma_k^-$ for dephasing environment [13]. Here $\sigma^\alpha$ ($\alpha = x, y, z$) are the usual Pauli spin operators. The master equation approach has been shown to be equivalent to the usual quantum operation approach for the description of decoherence in an open quantum system [21].

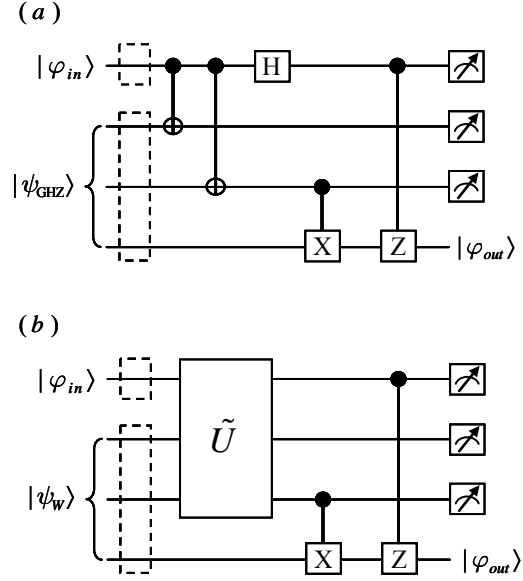

**Fig. 1.** Quantum gate circuits for teleportation in the presence of decoherence with the GHZ (a) and *W* (b) states. Here the top three lines belong to Alice, while the bottom one belongs to Bob. The "ammeter" symbol represents quantum measurement, and the dotted boxes denote decoherence channels. The unitary operator $\tilde{U}$ in (b) has the same form as that expressed in Eq. (3.1) of Ref. [17].

Without loss of generality, we consider as input the one-qubit state needs to be teleported in the form of $|\varphi_{in}\rangle = \cos(\theta/2)e^{i\phi/2}|0\rangle + \sin(\theta/2)e^{-i\phi/2}|1\rangle$, where $\theta \in [0, \pi]$ and $\phi \in [0, 2\pi]$ are the polar and azimuthal angles, respectively. Then if one adopts the three-qubit GHZ state $|\psi_{GHZ}\rangle = (|000\rangle + |111\rangle)/\sqrt{2}$ or the *W* state $|\psi_W\rangle = (\sqrt{2}|001\rangle + |010\rangle + |100\rangle)/2$ as quantum channel for teleportation [16,17] in decohering environment, the density operator for the output state has the form

$$\rho_{out} = \text{Tr}_{1,2,3}\{U_\alpha \varepsilon_1(\rho_{in}) \otimes \varepsilon_2(\rho_\alpha) U_\alpha^\dagger\}, \tag{2}$$

where $\rho_{in} = |\varphi_{in}\rangle\langle\varphi_{in}|$, $\rho_\alpha = |\psi_\alpha\rangle\langle\psi_\alpha|$ with $\alpha = \{GHZ, W\}$, and $\text{Tr}_{1,2,3}$ is a partial trace over the three qubits in possession of Alice. $\varepsilon_m$ ($m = 1, 2$) represents the quantum operations which transform the pure states $\rho_{in}$ and $\rho_\alpha$ to $\varepsilon_1(\rho_{in})$ and $\varepsilon_2(\rho_\alpha)$ due to the coupling of the system with its surrounding environment, and the explicit expressions for $\varepsilon_1(\rho_{in})$ and $\varepsilon_2(\rho_\alpha)$ can be derived by solving the master equation with corresponding initial conditions. Moreover, $U_\alpha$ are unitary operators which can be read directly from the teleportation circuits shown in Fig. 1.

To characterize the quality of the teleported state in decohering environment, we calculate the fidelity [21], defined as

$$F(\theta,\phi) = \langle \varphi_{in} | \rho_{out} | \varphi_{in} \rangle. \tag{3}$$

This quantity gives the information of how close the teleported state $\rho_{out}$ is to the unknown state $\rho_{in}$ to be teleported, i.e., they are equal when $F(\theta,\phi) = 1$ and orthogonal when $F(\theta,\phi) = 0$. Furthermore, by performing an average over all possible input states on the Bloch sphere one can get the average fidelity, which is explicitly expressed as

$$F_{av} = \frac{1}{4\pi} \int_0^{2\pi} d\phi \int_0^{\pi} d\theta \sin\theta F(\theta,\phi), \tag{4}$$

where $4\pi$ is the solid angle.

To begin with, we first explore the situation for which the state $\rho_{in}$ to be teleported is subject to different sources of decoherence while the channel state $\rho_\alpha$ ($\alpha = $ GHZ or $W$) is isolated perfectly from its surrounding environment, i.e., $\varepsilon_2(\rho_\alpha) \equiv \rho_\alpha$. For this special case, Bob can always get the same decohered state as $\rho_{out} = \varepsilon_1(\rho_{in})$ irrespective of the channel state (GHZ or $W$ state) shared with Alice. The explicit forms of $\varepsilon_1(\rho_{in})$ can be derived directly by solving the master equation expressed in Eq. (1) with the initial condition $\rho(0) = \rho_{in}$, the solutions can be summarized as $\rho^{11}(t) = \rho_{in}^{11} e^{-\gamma t}$, $\rho^{12,21}(t) = \rho_{in}^{12,21} e^{-\gamma t/2}$, and $\rho^{22}(t) = 1 - \rho_{in}^{11} e^{-\gamma t}$ for $\rho_{in}$ subject to zero temperature environment, $\rho^{11,22}(t) = [1 \pm (\rho_{in}^{11} - \rho_{in}^{22})e^{-2\gamma t}]/2$ and $\rho^{12,21}(t) = \rho_{in}^{12,21} e^{-\gamma t}$ for $\rho_{in}$ subject to infinite temperature environment, $\rho^{11,22}(t) = \rho_{in}^{11,22}$ and $\rho^{12,21}(t) = \rho_{in}^{12,21} e^{-\gamma t/2}$ for $\rho_{in}$ subject to dephasing environment. Combination of these results with Eq. (3), one can obtain the teleportation fidelities $F^{(\beta)}(\theta,\phi)$ ($\beta = z$, $i$ or $d$ indicates the state to be teleported is subject to zero temperature, infinite temperature or dephasing environment) as

$$F^{(z)}(\theta,\phi) = e^{-\gamma t} + \frac{1}{2}(e^{-\gamma t/2} - e^{-\gamma t})\sin^2\theta + (1-e^{-\gamma t})\sin^2\frac{\theta}{2},$$

$$F^{(i)}(\theta,\phi) = \frac{1}{2}[1 + e^{-2\gamma t} + (e^{-\gamma t} - e^{-2\gamma t})\sin^2\theta], \qquad (5)$$

$$F^{(d)}(\theta,\phi) = 1 + \frac{1}{2}(e^{-\gamma t/2} - 1)\sin^2\theta.$$

From the above three equations, one can see that all the teleportation fidelities are independent of the azimuthal angle $\phi$, while their values change with the variation of the polar angle $\theta$. This is different from that of the state $\rho_{in}$ subject to noise described by $L_1 = \sigma_1^x$ [14]. When subject to zero temperature environment, it is easy to show that the inequality $\partial F^{(z)}(\theta,\phi)/\partial\theta > 0$ holds in the whole time region, thus $F^{(z)}(\theta,\phi)$ always increases with increasing value of $\theta$. If $\theta = 0$ we attain its minimum $F^{(z)}_{\min} = e^{-\gamma t}$, and if $\theta = \pi$ we attain its maximum $F^{(z)}_{\max} = 1$. This indicates that only the excited state $|1\rangle$ can be perfectly teleported. When subject to infinite temperature environment, however, $F^{(i)}(\theta,\phi)$ decreases with the increase of $|\theta - \pi/2|$, and attains a certain minimum $F^{(i)}_{\min} = (1+e^{-2\gamma t})/2$ for $\theta = 0$ or $\pi$ and maximum $F^{(i)}_{\max} = (1+e^{-\gamma t})/2$ for $\theta = \pi/2$, thus no states can be teleported perfectly for this case. When subject to dephasing environment, the situation is completely reversed, i.e., $F^{(d)}(\theta,\phi)$ increases with increasing $|\theta - \pi/2|$, and attains its minimum $F^{(d)}_{\min} = (1+e^{-\gamma t/2})/2$ for $\theta = \pi/2$ and maximum $F^{(d)}_{\max} = 1$ for $\theta = 0$ or $\pi$. Now both the states $|0\rangle$ and $|1\rangle$ can be teleported perfectly.

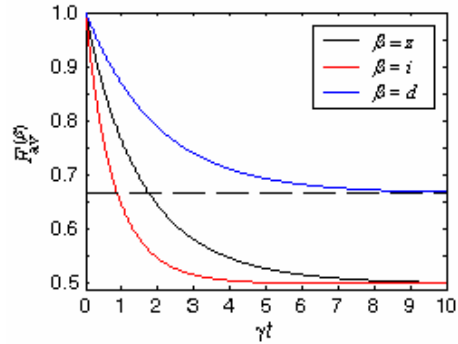

**Fig. 2.** (Color online) Average fidelity $F^{(\beta)}_{av}$ ($\beta = z, i, d$) versus $\gamma t$ for the case of the state to be teleported subject to different sources of decoherence.

Substituting Eq. (5) into Eq. (4) we get the average fidelity as

$$F_{\text{av}}^{(z)} = \frac{1}{2} + \frac{1}{3}e^{-\gamma t/2} + \frac{1}{6}e^{-\gamma t},$$

$$F_{\text{av}}^{(i)} = \frac{1}{2} + \frac{1}{3}e^{-\gamma t} + \frac{1}{6}e^{-2\gamma t}, \qquad (6)$$

$$F_{\text{av}}^{(d)} = \frac{2}{3} + \frac{1}{3}e^{-\gamma t/2}.$$

Plots of the above equations are shown in Fig. 2, from which one can observe that the presence of dephasing environment does not rule out the possibility for teleporting the one-qubit state with fidelity better than any classical communication protocol can offer since $F_{\text{av}}^{(d)} > 2/3$ in the whole time region [12]. For other two types of environments, however, there exists a critical rescaled time $\gamma t_c^{(\beta)}$ beyond which $F_{\text{av}}^{(\beta)} < 2/3$ ($\beta = z, i$). The critical rescaled time $\gamma t_c^{(\beta)}$ can be obtained analytically from Eq. (6) as $\gamma t_c^{(z)} = \ln(3 + 2\sqrt{2})$ and $\gamma t_c^{(i)} = \ln(1 + \sqrt{2})$, respectively.

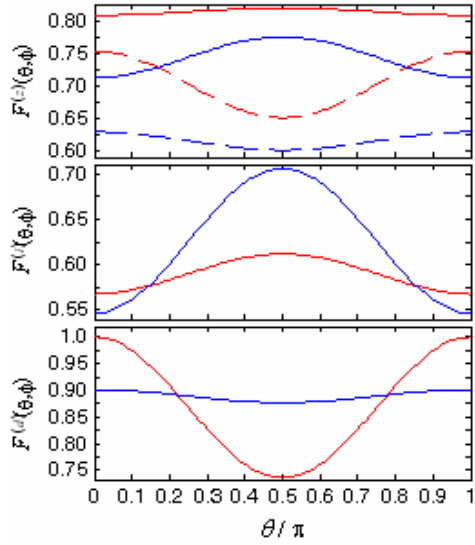

**Fig. 3.** (Color online) Fidelity $F^{(\beta)}(\theta, \phi)$ ($\beta = z, i, d$) versus $\theta/\pi$ for the case of the GHZ (the red lines) and W (the blue lines) states subject to different sources of decoherence. The topmost panel is plotted with $\gamma t = 0.3$ (solid lines) and $\gamma t = 0.8$ (dashed lines), while the middle and the bottommost panels are plotted with $\gamma t = 0.5$.

Now we turn our attention to the performance of quantum teleportation with the channel state shared and kept by Alice and Bob subject to different sources of decoherence while the state to be teleported is protected perfectly from decoherence, i.e., $\varepsilon_1(\rho_{\text{in}}) \equiv \rho_{\text{in}}$. We shall calculate the fidelity and average fidelity when the teleportation is implemented with $|\psi_{\text{GHZ}}\rangle$ and $|\psi_W\rangle$ as resources, and compare the robustness between them in terms of their teleportation capacity. The explicit expressions for $\varepsilon_2(\rho_\alpha)$ can be solved from Eq. (1) with the initial condition $\rho(0) = \rho_\alpha$ ($\alpha$ = GHZ, $W$). When $|\psi_{GHZ}\rangle$ is adopted as the quantum channel, the nonvanishing components of $\varepsilon_2(\rho_{GHZ})$ can be derived as: $\rho^{11} = e^{-3\gamma t}/2$, $\rho^{18,81} = e^{-3\gamma t/2}/2$, $\rho^{22,33,55} = (e^{-2\gamma t} - e^{-3\gamma t})/2$, $\rho^{44,66,77} = e^{-\gamma t}/2 - e^{-2\gamma t} + e^{-3\gamma t}/2$, and $\rho^{88} = 1 - 3e^{-\gamma t}/2 + 3e^{-2\gamma t}/2 - e^{-3\gamma t}/2$ for the situation of zero temperature environment, $\rho^{11,88} = (1+3e^{-4\gamma t})/8$, $\rho^{18,81} = e^{-3\gamma t}/2$, and $\rho^{22-77} = (1-e^{-4\gamma t})/8$ for the situation of infinite temperature environment, $\rho^{11,88} = 1/2$ and $\rho^{18,81} = e^{-3\gamma t/2}/2$ for the situation of dephasing environment. These results, together with Eqs. (2) and (3) yield

$$F^{(z)}(\theta,\phi) = 1 - e^{-\gamma t} + e^{-2\gamma t} - \frac{1}{2}(1 - e^{-3\gamma t/2} - 2e^{-\gamma t} + 2e^{-2\gamma t})\sin^2\theta,$$
$$F^{(i)}(\theta,\phi) = \frac{1}{2}[1 + e^{-4\gamma t} + (e^{-3\gamma t} - e^{-4\gamma t})\sin^2\theta], \quad (7)$$
$$F^{(d)}(\theta,\phi) = 1 - \frac{1}{2}(1 - e^{-3\gamma t/2})\sin^2\theta.$$

The plots of Eq. (7) with fixed rescaled time $\gamma t$ are displayed in Fig. 3 as red lines. For zero temperature environment, since $\partial F^{(z)}(\theta,\phi)/\partial\theta = (e^{-3\gamma t/2} + 2e^{-\gamma t} - 2e^{-2\gamma t} - 1)\sin\theta\cos\theta$, where $e^{-3\gamma t/2} + 2e^{-\gamma t} - 2e^{-2\gamma t} - 1 > 0$ if $\gamma t < \gamma t_c^{(z)} \simeq 0.3739$, and $e^{-3\gamma t/2} + 2e^{-\gamma t} - 2e^{-2\gamma t} - 1 < 0$ if $\gamma t > \gamma t_c^{(z)}$, the teleportation fidelity $F^{(z)}(\theta,\phi)$ is a decreasing function of $|\theta - \pi/2|$ when $\gamma t < \gamma t_c^{(z)}$, and an increasing function of $|\theta - \pi/2|$ when $\gamma t > \gamma t_c^{(z)}$. For infinite temperature environment, however, $\partial F^{(i)}(\theta,\phi)/\partial\theta = (e^{-3\gamma t} - e^{-4\gamma t})\sin\theta\cos\theta$, thus the teleportation fidelity $F^{(i)}(\theta,\phi)$ is always a decreasing function of $|\theta - \pi/2|$ in the whole time region. Finally, for the dephasing environment, since $\partial F^{(d)}(\theta,\phi)/\partial\theta = (e^{-3\gamma t/2} - 1)\sin\theta\cos\theta$, the situation becomes completely reversed, i.e., $F^{(d)}(\theta,\phi)$ is always an increasing function of $|\theta - \pi/2|$ in the whole time region. Moreover, different from those of the zero and infinite temperature environments where no states can be teleported perfectly, here both the states $|0\rangle$ and $|1\rangle$ can be teleported with unity fidelity.

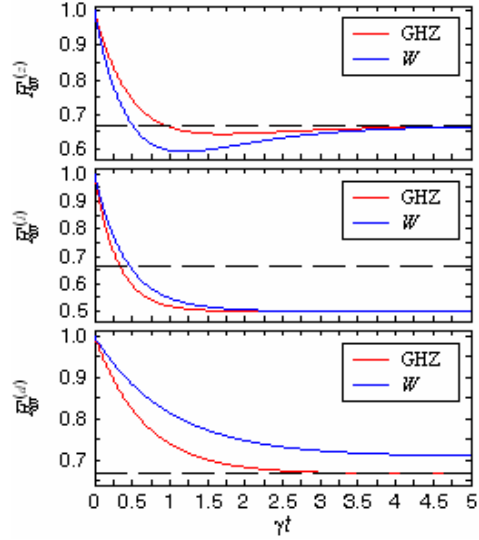

**Fig. 4.** (Color online) Average fidelity $F_{av}^{(\beta)}$ ($\beta = z, i, d$) versus $\gamma t$ for the case of the GHZ and $W$ states used as quantum channel subject to different sources of decoherence. Here the dashed lines at $F_{av}^{(\beta)} = 2/3$ show the highest fidelity for classical transmission of a quantum state.

By performing an average over all possible input states on the Bloch sphere, one can further obtain the average fidelity as

$$F_{av}^{(z)} = \frac{2}{3} - \frac{1}{3}e^{-\gamma t} + \frac{1}{3}e^{-2\gamma t} + \frac{1}{3}e^{-3\gamma t/2},$$
$$F_{av}^{(i)} = \frac{1}{2} + \frac{1}{3}e^{-3\gamma t} + \frac{1}{6}e^{-4\gamma t}, \qquad (8)$$
$$F_{av}^{(d)} = \frac{2}{3} + \frac{1}{3}e^{-3\gamma t/2}.$$

Plots of the above equations are shown in Fig. 4 as red solid lines. Clearly, the average fidelity for zero temperature environment initially decreases with increasing value of the rescaled time $\gamma t$ and arrives at a certain minimum which is smaller than $2/3$, and then it begins to increase with the increase of $\gamma t$ and finally approaches to its asymptotic value $2/3$. This is different from that of the GHZ state subject to noise modeled by the Pauli spin operators [19]. For other two sources of decoherence, however, the average fidelities always decay exponentially with increasing $\gamma t$. Moreover, from Fig. 4 one can also observe that for dephasing environment, the quantum protocol always outperform those of classical ones since $F_{av}^{(d)}$ approaches to its limiting value $2/3$ only when $\gamma t_c^{(d)} \to \infty$ [12]. For zero and infinite temperature environments, however, the performance of quantum teleportation protocol becomes worse than the best possible score when Alice and Bob communicate with each other only via the classical channel [12] after a critical rescaled time $\gamma t_c^{(\beta)}$

- 8 -

($\beta = z, i$). For zero temperature environment, $\gamma t_c^{(z)}$ can be obtained analytically from the first expression in Eq. (8) as $\gamma t_c^{(z)} = \ln[(3+\sqrt{5})/2]$, while for infinite temperature environment, $\gamma t_c^{(i)}$ can only be obtained numerically, the result is $\gamma t_c^{(i)} \simeq 0.3331$. Moreover, by comparing the above results with Fig. 1 in Ref. [13], one can also notice that at the critical point $\gamma t_c^{(\beta)}$ ($\beta = z, i$), the multipartite concurrence $C_3^{(\beta)}(t_c)$ remains nonzero. This indicates that in order to teleport $|\varphi_{in}\rangle$ with fidelity larger than $2/3$ the quantum channel should possess a critical value of minimum multipartite entanglement. For the dephasing environment, however, a nonzero critical value of minimum entanglement is not necessary because we have $\gamma t_c^{(d)} \to \infty$ and $C_3^{(d)}(t_c) \to 0$.

When $|\psi_W\rangle$ is used as the quantum channel, setting $u = e^{-\gamma t} - e^{-2\gamma t}$, $v_\pm = e^{-2\gamma t} \pm e^{-4\gamma t}$, and $w_\pm = 1 \pm e^{-6\gamma t}$, then by solving Eq. (1), the nonvanishing components of $\varepsilon_2(\rho_W)$ can be derived explicitly as: $\rho^{22} = \sqrt{2}\rho^{23,25} = 2\rho^{33,55,35} = e^{-2\gamma t}/2$, $\rho^{44,66} = 3\rho^{46} = 3\sqrt{2}\rho^{47,67}/2 = 3\rho^{77}/2 = 3u/4$, and $\rho^{88} = 1 - 2e^{-\gamma t} + e^{-2\gamma t}$ with $\rho^{ij} = \rho^{ji}$ for zero temperature environment, $\rho^{11,77} = (w_- \pm v_-)/8$, $\rho^{22,88} = (w_+ \pm v_+)/8$, $\rho^{23,25} = \sqrt{2}v_+/8$, $\rho^{47,67} = \sqrt{2}v_-/8$, $\rho^{35,46} = v_\pm/8$, $\rho^{33,55} = w_+/8$, and $\rho^{44,66} = w_-/8$ with $\rho^{ij} = \rho^{ji}$ for the infinite temperature environment, $\rho^{22} = 2\rho^{33,55} = 1/2$ and $\rho^{23,25} = \sqrt{2}\rho^{35} = \sqrt{2}e^{-\gamma t}/4$ with $\rho^{ij} = \rho^{ji}$ for the dephasing environment. Combination of the above results with Eqs. (2) and (3), one can obtain the teleportation fidelity as

$$F^{(z)}(\theta,\phi) = 1 - \frac{3}{2}e^{-\gamma t} + \frac{3}{2}e^{-2\gamma t} - \frac{1}{2}(1 - 3e^{-\gamma t} + 2e^{-2\gamma t})\sin^2\theta,$$
$$F^{(i)}(\theta,\phi) = \frac{1}{4}(2 + e^{-4\gamma t} + e^{-6\gamma t}) + \frac{1}{2}(e^{-2\gamma t} - e^{-6\gamma t})\sin^2\theta, \qquad (9)$$
$$F^{(d)}(\theta,\phi) = \frac{1}{4}(3 + e^{-\gamma t}) - \frac{1}{16}(1 - e^{-\gamma t})\sin^2\theta.$$

The fidelities expressed in Eq. (9) versus $\theta/\pi$ for fixed rescaled time $\gamma t$ are plotted in Fig. 3 as blue lines, from which one can see that $F^{(\beta)}(\theta,\phi)$ ($\beta = z, i, d$) here show similar behaviors as those with $|\psi_{GHZ}\rangle$ as the quantum channel. This can be understood directly from Eq. (9) as follows. For zero temperature environment, $\partial F^{(z)}(\theta,\phi)/\partial\theta = (3e^{-\gamma t} - 2e^{-2\gamma t} - 1)\sin\theta\cos\theta$. It can be obtained analytically that $3e^{-\gamma t} - 2e^{-2\gamma t} - 1 > 0$ if $\gamma t < \ln 2$, which induces the decay of $F^{(z)}(\theta,\phi)$ with increasing value of $|\theta - \pi/2|$, while $3e^{-\gamma t} - 2e^{-2\gamma t} - 1 < 0$ if $\gamma t > \ln 2$, which induces the enhancement of $F^{(z)}(\theta,\phi)$ with increasing value of $|\theta - \pi/2|$. Similarly, for other two sources of surrounding environments we have $\partial F^{(i)}(\theta,\phi)/\partial\theta = (e^{-2\gamma t} - e^{-6\gamma t})\sin\theta\cos\theta$ and $\partial F^{(d)}(\theta,\phi)/\partial\theta = [(e^{-\gamma t} - 1)\sin\theta\cos\theta]/8$, thus the behaviors displayed in Fig. 3 can also be understood analogously.

Substituting Eq. (9) into Eq. (4), we obtain the average fidelity as

$$F_{av}^{(z)} = \frac{2}{3} - \frac{1}{2}e^{-\gamma t} + \frac{5}{6}e^{-2\gamma t},$$
$$F_{av}^{(i)} = \frac{1}{2} + \frac{1}{3}e^{-2\gamma t} + \frac{1}{4}e^{-4\gamma t} - \frac{1}{12}e^{-6\gamma t}, \quad (10)$$
$$F_{av}^{(d)} = \frac{17}{24} + \frac{7}{24}e^{-\gamma t}.$$

The $\gamma t$ dependence of $F_{av}^{(\beta)}$ ($\beta = z, i, d$) are plotted in Fig. 4 as blue solid lines. Similar to that with $|\psi_{GHZ}\rangle$ as the resource, the average fidelity $F_{av}^{(z)}$ here still does not behave as a monotonous function of $\gamma t$. $F_{av}^{(z)}$ first decays to a minimum of $71/120$, and then it begins to increase and finally approaches to the classical limiting value of $2/3$. Moreover, there exists a rescaled critical time $\gamma t_c^{(z)} = \ln(5/3)$ beyond which the performance of quantum teleportation in zero temperature environment is worse than the best possible score that the classical communication protocol can offer [12]. For the situation of infinite temperature or dephasing environment, however, both $F_{av}^{(i)}$ and $F_{av}^{(d)}$ decay monoexponentially with increasing $\gamma t$ in the whole time region. $F_{av}^{(i)}$ becomes smaller than $2/3$ after $\gamma t > \gamma t_c^{(i)} \simeq 0.4615$, while the asymptotic value of $F_{av}^{(d)}$ is $17/24$, which is slightly larger than $2/3$.

Finally, we would like to make a comparison of the robustness between the GHZ and W states in terms of their teleportation capacity. As indicated by the red and blue lines shown in Fig. 3, when subject to zero temperature environment, $F^{(z)}(\theta,\phi)$ for the GHZ state is larger than that for the W state in the entire range of $\theta$. This contributes to the fact that $F_{av}^{(z)}$ for the GHZ state is larger than that for the *W* state (see the top panel of Fig. 4), which indicates that the GHZ state is always robust compared to the *W* state under the influence of this type of channel-environment coupling. When subject to infinite temperature or dephasing environment, however, the situation is completely reversed. Now the *W* state is more robust than the GHZ state (see the bottom two panels of Fig. 4). This can be understood from Fig. 3, where $F^{(i)}(\theta,\phi)$ for the *W* state is larger than that for the GHZ state in almost every range of $\theta$ (approximately $0.1418 < \theta/\pi < 0.8582$) except for the small boundary region, and $F^{(d)}(\theta,\phi)$ for the *W* state is also larger in the middle $\theta$ region (approximately $0.2216 < \theta/\pi < 0.7784$).

To summarize, we have investigated quantum teleportation in different sources of surrounding environments, and focused on the comparison of the robustness between the GHZ and *W* states in terms of their teleportation capacity. Our results revealed that when the state to be teleported or the channel state shared by Alice and Bob is subject to decoherence, only in dephasing environment can the teleportation fidelity $F_{av}^{(d)}$ is always larger than the best possible score $2/3$ of classical communication, while for the zero and infinite temperature environments, $F_{av}^{(z)}$ and $F_{av}^{(i)}$ become smaller than $2/3$ after a finite time. Moreover, the robustness of GHZ and *W* states is dependent on the environment types. When subject to zero temperature environment, the GHZ state is always robust than the *W* state, while for infinite temperature and dephasing environments, the situation becomes completely reversed, i.e., the *W* state is robust compared to the GHZ state.

## Acknowledgments

This work was supported by the NSF of Shaanxi Province under Grant Nos. 2010JM1011 and 2009JQ8006, the Specialized Research Program of Education Department of Shaanxi Provincial Government under Grant Nos. 2010JK843 and 2010JK828, and the Youth Foundation of XUPT under Grant No. ZL2010-32.